\documentclass{epl}
\usepackage{amssymb}
\usepackage{graphics}
\usepackage{graphicx}

\title{The distribution function of a semiflexible polymer and random walks
with constraints}
\author{Semjon Stepanow\inst{1} \and Gunter M. Sch\"utz\inst{2}}
\institute{
 \inst{1} Martin-Luther-Universit\"at Halle, Fachbereich Physik, D-06099 Halle, Germany\\
 \inst{2} Institut f\"ur Festk\"orperforschung, Forschungszentrum J\"ulich, D-52425 J\"ulich, Germany 
}
 {Fluctuation Phenomena, random processes,noise, and Brownian motion}

\begin{document}

\maketitle

\begin{abstract}
In studying the end-to-end distribution function $G(r,N)$ of a worm like
chain by using the propagator method we have established that the
combinatorial problem of counting the paths contributing to $G(r,N)$ can be
mapped onto the problem of random walks with constraints, which is closely
related to the representation theory of the Temperley-Lieb algebra. By using
this mapping we derive an exact expression of the Fourier-Laplace transform
of the distribution function, $G(k,p)$, as a matrix element of an inverse of
an infinite rank matrix. Using this result we also derived a recursion
relation permitting to compute $G(k,p)$ directly. We present the results of
the computation of $G(k,N)$ and its moments. The moments $<r^{2n}>$ of $%
G(r,N)$ can be calculated \emph{exactly} by calculating the (1,1) matrix
element of $2n$-th power of a truncated matrix of rank $n+1$.
\end{abstract}

The theory of flexible polymers is now understood \cite{edwards65}-\cite%
{doi-edwards-book}. Many polymer molecules have internal stiffness and
cannot be modeled by the model of flexible molecules developed by Edwards.
This is especially true for several important biopolymers such as actin,
DNA, and microtubules \cite{sackmann94}. Models of semiflexible polymers
have also applications in different topics besides polymer physics \cite%
{liverpool-edwards}. If the chain length decreases, the chain stiffness
becomes an important factor. A quantitative measure for the stiffness of the
polymer is the persistence length $l_{p}$, which is the correlation length
for the tangent-tangent correlation function along the polymer. Polymers
with contour length $L$ much larger than $l_{p}$ are flexible and are
described by using the tools of quantum mechanics and quantum field theory 
\cite{edwards65}-\cite{doi-edwards-book}. The standard coarse-graining model
of a wormlike or a semiflexible polymer was proposed by Kratky and Porod 
\cite{kratky/porod49}. A few first moments of $G(r,N)$ were computed in \cite%
{hermans/ullman52}-\cite{saitoetal67}. The literature on the computation of $%
G(r,N)$ and its moments can be found in the book by Yamakawa \cite{yamakawa}. 
For recent work see \cite{wilhelm/frey96}-\cite{samuel-sinha02}.

In this Letter we will study the problem of computation of the distribution
function $G(r,N)$ of the end-to-end distance of a semiflexible polymer,
which is described by Kratky-Porod model, by using the analogy of the worm
like chain with the quantum rigid rotator in an external homogeneous field 
\cite{descloizeaux},\cite{zinn-justin}
within the quantum mechanical propagator method \cite{feynman-hibbs}. The relation to the rigid 
rotator was used in the recent work \cite{samuel-sinha02}. 
Relating the combinatorics of counting the paths contributing to $G(r,N)$ to
random walks with constraints and the Temperley-Lieb algebra \cite%
{temperley-lieb71}, we have derived an exact formula for the Fourier-Laplace
transform of the end-to-end distribution function, $G(k,p)$, as an inverse
of an infinite rank matrix of a very simple structure, given by Eqs.(\ref{w6}%
-\ref{w7}). Using a truncated matrix of rank $n+1$ one obtains $n$ moments
of $r^{2}$ \emph{exactly} from the (1,1) element of the $2n$-th power of
this matrix, i.e., by applying this matrix $2n$ times to a canonical basis
vector. Moreoever, this enables one to compute the distribution function,
which describes $n$ moments of $r^{2}$ \emph{exactly} and all higher moments
approximately. The truncation corresponds to taking into account the
intermediate states with the quantum number of the angular momentum $l\leq n$
(see below).

The Fourier transform of the distribution function of the end-to-end polymer
distance of the Kratky-Porod model \cite{kratky/porod49} $G(\mathbf{k}%
,L)=\int d^{3}R\exp (-i\mathbf{k(\mathbf{R}-\mathbf{R}_{0})})G(\mathbf{R}-%
\mathbf{R}_{0},L)$ is given by the path integral as follows 
\begin{equation}
G(\mathbf{k},L)=\int D\mathbf{t}(s)\prod\limits_{s}\delta (\mathbf{t}%
(s)^{2}-1)\exp (-i\mathbf{k}\int_{0}^{L}ds\mathbf{t}(s)-\frac{l_{p}}{2}%
\int_{0}^{L}ds(\frac{\partial \mathbf{t}(s)}{\partial s})^{2}),  \label{w1}
\end{equation}%
where $l_{p}$ is the persistence length, and $\mathbf{t}(s)$ is the tangent
vector at the point $s$ along the contour of the polymer. The product over $%
s $ in Eq.(\ref{w1}) takes into account that the polymer chain is locally
inextensible. If the polymer interacts with an external field, the potential
energy $-\int_{0}^{L}dsV(\mathbf{r}(s))$ should be added in the exponential
of Eq.(\ref{w1}). Notice that the representation of $G(k,L)$ by using the
cumulants of the moments of the end-to-end distribution function $G(\mathbf{k%
},L)=\exp (-\sum_{m=1}^{\infty }\int_{0}^{L}ds\mu _{2m}(s)(\mathbf{k}%
^{2})^{m})$ permits to write the following differential equation for $G(%
\mathbf{R},L)$, $\partial \ G(\mathbf{R},L)/\partial L-\sum_{m=1}^{\infty
}(-1)^{m+1}\mu _{2m}(L)\Delta _{\mathbf{R}}^{m}G=0$, where $\Delta =\nabla
^{2}$ is the Laplace operator. The latter with the term associated with the
external potential was used in \cite{stepanow/JCP} to describe the
localization of a semiflexible polymer onto interfaces and surfaces.

The path integral (\ref{w1}) (without the term depending on $k$) corresponds
to the diffusion of a particle on unit sphere, $\left\vert \mathbf{t}%
(s)\right\vert =1$. This problem is equivalent to the Euclidean rigid
quantum rotator \cite{descloizeaux},\cite{zinn-justin}. The Green's function
of the rigid rotator or a particle on unit sphere fulfills the following
equation%
\begin{equation}
\frac{\partial }{\partial L}P_{0}(\theta ,\varphi ,L;\theta _{0},\varphi
_{0},0)-\frac{1}{2l_{p}}\nabla _{\theta ,\varphi }^{2}P_{0}=\delta (L)\delta
(\Omega -\Omega _{0}),  \label{w2}
\end{equation}%
where $\Omega $ is the spheric angle associated with angles $\theta $, and $%
\varphi $, and $\delta (\Omega -\Omega _{0})$ is a two dimensional delta
function. Henceforth instead of the contour length $L$ we will use the
number of statistical segments $N=L/l_{p}$. Notice that in Quantum Mechanics 
$N$ corresponds to the imaginary time $it$. The solution for $P_{0}(\theta
,\varphi ,N;\theta _{0},\varphi _{0},0)$ is 
\begin{equation}
P_{0}(\theta ,\varphi ,N;\theta _{0},\varphi _{0},0)=\sum_{l,m}\exp (-\frac{%
l(l+1)N}{2})Y_{lm}(\theta ,\varphi )Y_{lm}^{\ast }(\theta _{0},\varphi _{0}),
\label{w3}
\end{equation}%
where $Y_{lm}(\theta ,\varphi )$ are the spherical harmonics. At a given $l$%
, $m$ takes the values $-l$, $-l+1$, $...$, $l$. The quantity $P_{0}(\theta
,\varphi ,N;\theta _{0},\varphi _{0},0)$ is given by Eq.(\ref{w1}) with $%
\mathbf{k}=0$ and with the following boundary conditions for the path $%
\mathbf{t}(s)$ ($0\leq s\leq N$): $\mathbf{t}(N)\equiv (\theta ,\varphi )$,
and $\mathbf{t}(0)\equiv (\theta _{0},\varphi _{0})$.

We now will consider the Green's function $P(\theta ,\varphi ,N;\theta
_{0},\varphi _{0},0)$ associated with Eq.(\ref{w1}). The differential
equation for $P$ is%
\begin{equation}
\frac{\partial }{\partial N}P(\theta ,\varphi ,N;\theta _{0},\varphi _{0},0)-%
\frac{1}{2}\nabla _{\theta ,\varphi }^{2}P+U(\Omega )P=\delta (N)\delta
(\Omega -\Omega _{0}),  \label{w4}
\end{equation}%
where $U(\Omega )=i\mathbf{k\Omega =}ik\cos (\theta )$, and $k$ is measured
in units of $l_{p}^{-1}$. As it is well-known from the propagator method 
\cite{feynman-hibbs} the latter can be written as an integral equation as
follows%
\begin{equation}
P(\Omega ,N;\Omega _{0},0)=P_{0}(\Omega ,N;\Omega _{0},0)-\int_{0}^{N}ds\int
d\Omega ^{\prime }P_{0}(\Omega ,N;\Omega ^{\prime },s)U(\Omega ^{\prime
})P(\Omega ^{\prime },s;\Omega _{0},0).  \label{w5}
\end{equation}%
Eqs.(\ref{w3}-\ref{w5}) describes also the Euclidean rigid quantum rotator
in an external field. The iteration of the latter generates the perturbation
expansion of $P$ in powers of the potential $U(\Omega )$. The Fourier
transform of the distribution function of the end-to-end distance of the
polymer is obtained from $P(\Omega ,N;\Omega _{0},0)$ by integrating the
latter over $\Omega $ and $\Omega _{0}$: $G(k,N)=\int d\Omega \int d\Omega
_{0}P(\Omega ,N;\Omega _{0},0)$. The factor $4\pi $ is extracted from $%
P(\Omega ,N;\Omega _{0},0)$. Eq.(\ref{w5}) with fixed $\Omega $ and $\Omega
_{0}$ is the distribution function of the end-to end distance with fixed
tangents of the polymer ends. The coefficient in front of ($k^{2})^{n}$ of
the expansion of $G(k,p)$ in powers of $k^{2}$, multiplied by the factor $%
(-1)^{n}\Gamma (2n+2)$ is the Laplace transform of the moment $<r^{2n}>$ of
the end-to-end distribution function. Due to the folding character of the
integrations over the contour lengths in (\ref{w5}), the Laplace transform
of (\ref{w3}) with respect to $N$ permits to get rid of integrations over
the contour length. Similar to the propagator method used elsewhere \cite%
{feynman-hibbs} each order of the perturbation expansion can be represented
by graphs, examples of which are schematically shown in Fig.1. 
\begin{figure}[tbph]
\onefigure{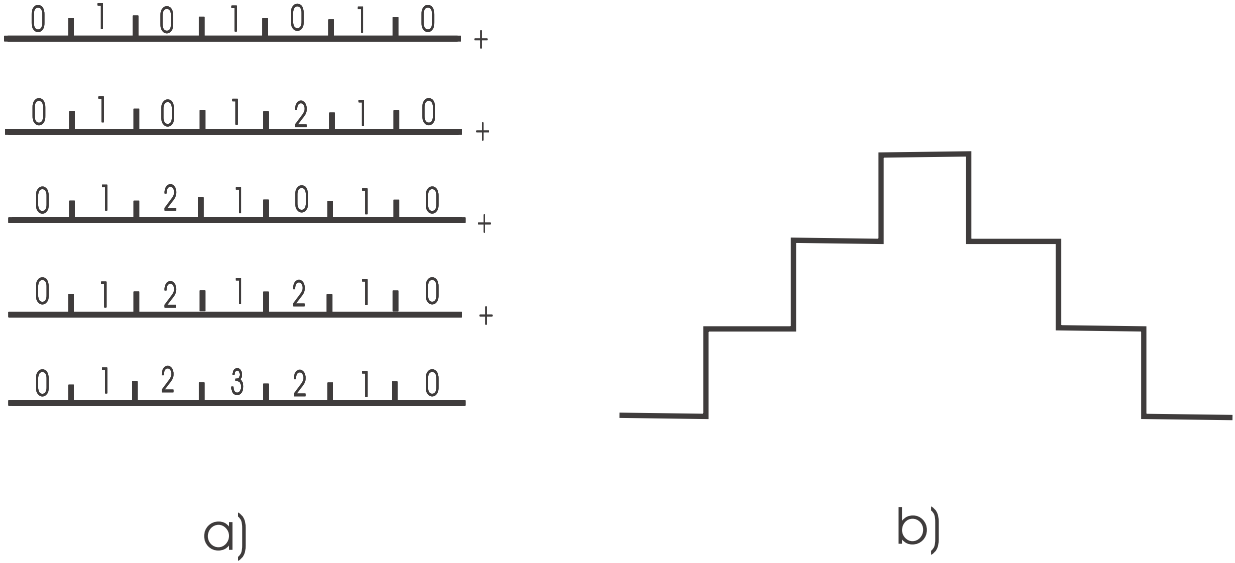}
\caption{a) The $6th$ order of the perturbation expansion of the end-to-end
distribution function. b) Random walk with constraint associated with the
last line in a).}
\label{Fig.1}
\end{figure}
Each piece of the continuous line between two consecutive vertices are
associated with the Laplace transform of the bare Green's function given by
Eq.(\ref{w3}). The integral associated with each vertex is $\left\langle
\cos \theta \right\rangle _{l^{\prime },l;m^{\prime },m}=\int d\Omega
Y_{l^{\prime },m^{\prime }}^{\ast }(\theta ,\varphi )\cos \theta
Y_{l,m}(\theta ,\varphi )$. Using the property of the spherical functions $%
\cos \theta Y_{l,m}=\sqrt{((l+1)^{2}-m^{2})/(4(l+1)^{2}-1)}Y_{l+1,m}+\sqrt{%
(l^{2}-m^{2})/(4l^{2}-1)}Y_{l-1,m}$ gives $\left\langle \cos \theta
\right\rangle _{l^{\prime },l;m^{\prime },m}=\sqrt{%
((l+1)^{2}-m^{2})/(4(l+1)^{2}-1)}\delta _{l^{\prime },l+1}\delta _{m^{\prime
},m}+\sqrt{(l^{2}-m^{2})/(4l^{2}-1)}\delta _{l^{\prime },l-1}\delta
_{m^{\prime },m}$.

Due to the integrations over $\Omega $ and $\Omega _{0}$ the quantum numbers 
$l$ and $m$ associated with the transition probabilities (amplitudes in the
quantum mechanical counterpart of the problem) at graph ends are equal to
zero. The averages over $\cos \theta $ in each vertex gives that the quantum
numbers $l$ on both sides of a vertex differ by one. The weights associated
with the vertices are given by $w_{l}=\left\langle \cos \theta \right\rangle
_{l,l-1}=\left\langle \cos \theta \right\rangle _{l-1,l}=l/\sqrt{4l^{2}-1}$.
The Laplace transform of the weights associated with each piece of the
continuous line between two vertices, which are given by the exponential
factor in Eq.(\ref{w3}), are given by $e_{l}=1/(p+l(l+1)/2)$, where $p$ is
the Laplace conjugate to $N$. The quantum number $m$ is conserved along the
path and is zero due to the integration over the angles at the ends. Thus,
the problem of counting different paths contributing to the transition
probability $\int d\Omega \int d\Omega _{0}P(\Omega ,N;\Omega _{0},0)$ can
be formulated as a problem of a random walk consisting of odd number of
steps with the constraint that the step heights are equal to one. The
example of such a random walk is given in Fig.1b. The computation of the
coefficients in front of $(k^{2})^{n}$ of the expansion of $G(k,p)$ in
powers of $k^{2}$ is thus related to the enumeration of configurations of
such a random walk consisting of $2n+1$ steps. This kind of the problem is
closely related to the calculation of the dimensions of the irreducible
representations of the Temperley-Lieb algebra \cite{temperley-lieb71}. One
can write the expansion of $G(k,p)$ as follows%
\begin{equation}
G(k,p)=\left[ d+\sum\limits_{n=1}^{\infty }(-k^{2})^{n}A^{2n}\right] _{1,1}=%
\frac{1}{p}\left[ I+k^{2}A^{2}\right] _{1,1}^{-1},  \label{w6}
\end{equation}%
where the last term is the result of the summation of the series, $A=d.m$ is
the product of the infinite rank matrices $d$ and $m$, which are defined as
follows $d_{i,j}=e_{i-1}\delta _{i,j}$ and $m_{i,j}=w_{i-1}\delta
_{i,j+1}+w_{i}\delta _{i,j-1}$ ($i$,$j=1$, $2$, $...$). Approximating the
last term in Eq.(\ref{w6}) by the matrices of finite rank $n$ gives an
approximate formula for the distribution function, $G_{2n-2}(k,p)$, which
gives exactly the first $2n-2$ moments and approximate expressions for
higher moments where the intermediate states with $l=0,1,...,n-1$ are taken
into account. No matrix inversion is required for the exact Laplace
transforms of the moments which may be calculated directly by the Fourier
transform of (\ref{w6}). From the expansion in $k$ one realizes that only
the (1,1)-element of the $2n$-power of $A$ needs to be calculated for $%
<r^{2n}>$. Because of the simple band structure of $A$ this is obtained
exactly from a truncated matrix $\tilde{A}$ of rank $n+1$.

The distribution function $G_{2n}(k,p)$ ($2n$ is the number of the exact
first moments contained in $G_{2n}(k,p)$) can be written as 
\begin{equation}
G_{2n}(k,p)=\frac{1}{p}\frac{1+f_{2}+f_{3}+...+f_{n}}{%
1+f_{1}+f_{2}+f_{3}+...+f_{n}},  \label{w7}
\end{equation}%
where $f_{k}$, which are defined for each $n$, obey the following recursion
relation $f_{k}=u_{k}(1+\sum_{l=k+2}^{n}f_{l})$, and $%
u_{i}=k^{2}e_{i-1}w_{i}^{2}e_{i}$ . Eqs.(\ref{w6}-\ref{w7}) are exact and
are the main result of the present Letter. The off diagonal matrix elements
of the matrix $\left[ I+k^{2}A^{2}\right] ^{-1}$ give the transition
probabilities from the initial state ($l_{a}$, $m_{a}=0$) to the final state
($l_{b}$, $m_{b}=0$). The computation of the distribution function of the
end-to-end distance with fixed tangents of the end vectors demands
consideration of paths with arbitrary $l_{a}$, $m_{a}$ and $l_{b}$, $m_{b}$.
The consideration of the transition probability with nonzero number $m$
demands the change of the definition of the weights $w_{l}$. The simple band
structure of the matrix, however, remains unchanged.

We have computed $G_{m}(k,p)$ explicitly for $m\leq 26$. The distribution
function, which takes into account exact $6$ moments, is given by%
\[
G_{6}(k,p)=\frac{5}{3}\frac{39\,k^{2}+11\,k^{2}p+21\,{p}^{3}+210\,{p}%
^{2}+378+567\,p}{35\,{p}^{4}+350\,{p}^{3}+30\,k^{2}{p}^{2}+945\,{p}%
^{2}+630\,p+170\,k^{2}p+3\,k^{4}+210\,k^{2}}. 
\]%
As an example we give below the $10th$ moment 
\begin{eqnarray*}
<(r^2)^{5}>&=&-{\frac{256}{540225}}\,N{e^{-10\,N}}-{\frac{1024}{3781575}}\,{%
e^{-10\,N}}+{\frac{1408}{50625}}\,N^{2}{e^{-6\,N}}+{\frac{15616}{455625}}\,N{%
e^{-6\,N}}- \\
&&{\frac{15872}{20503125}}\,{e^{-6\,N}}-{\frac{7040}{3969}}\,N^{3}{e^{-3\,N}}%
+{\frac{14080}{9261}}\,N^{2}{e^{-3\,N}}+{\frac{1303040}{21609}}\,N{e^{-3\,N}}%
+ \\
&&{\frac{1285273600}{12252303}}\,{e^{-3\,N}}+{\frac{22176}{25}}\,N^{4}{e^{-N}%
}+{\frac{4549248}{125}}\,N^{3}{e^{-N}}+{\frac{2338481024}{4375}}\,N^{2}{%
e^{-N}}+ \\
&&{\frac{973888866304}{275625}}\,N{e^{-N}}+{\frac{88900199815424}{9646875}}\,%
{e^{-N}}+{\frac{256}{56260575}}\,{e^{-15\,N}}+{\frac{12320}{9}}\,N^{5}- \\
&&{\frac{726880}{27}}\,N^{4}+{\frac{21670528}{81}}\,N^{3}-{\frac{651701248}{%
405}}\,N^{2}+{\frac{34520038912}{6075}}\,N- \\
&&{\frac{167953309696}{18225}.}
\end{eqnarray*}%
All $26$ moments we have computed are obtained in a form similar to $%
<(r^2)^{5}> $ with numerical coefficients given as fractional numbers. All
moments fulfil the expected limiting behaviour for small and large $N$.

Performing the inverse Laplace transform of $G(k,p)$ by using Maple enables
one to plot $G(k,N)$. In the limit of small $N$ the distribution function $%
G_{m}(r,N)$ approaches the rod limit $G_{rod}(r,N)=1/(4\pi N^{2})\delta
(r-N) $, which has the Fourier transform $G_{rod}(k,N)=\sin (kN)/kN$. In
order to illustrate the accuracy of the truncated dustribution function a
comparison of $G(k,N)$ with the rod limit is shown in Fig.2. 
\begin{figure}[tbph]
\onefigure{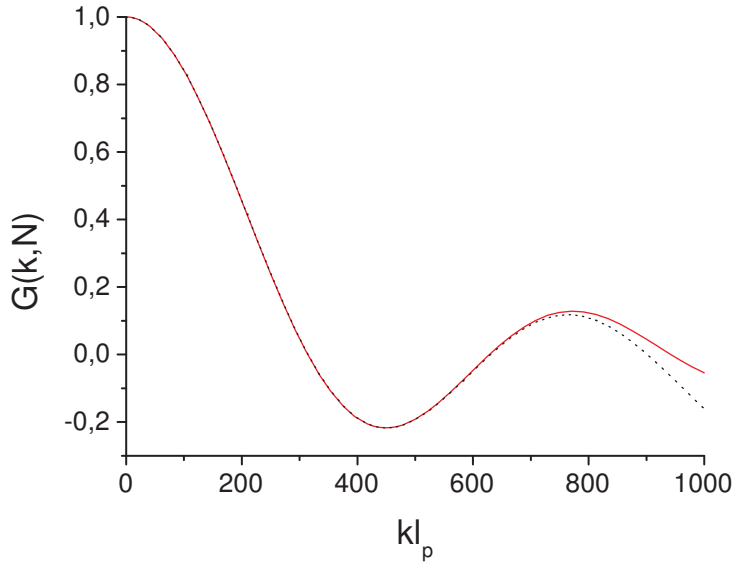}
\caption{Computation of $G(k,N)$ for $N=0.01$: Continuous: $G_{rod}(k,N)$,
dots: $G_{10}(k,N)$}
\label{Fig.2}
\end{figure}
The increase of $m$ extends the interval of $k$ of the agreement between $%
G_{m}(k,N)$ and $G_{rod}(k,N)$. However, for large $m$ the Maple computation
of the inverse Laplace transform of $G(k,p)$ becomes unstable for small $k$
at very small $N$.

The inverse Laplace transform of $G(k,p)/p^{2}$ multiplied with the factor $%
2/N$ gives the structure factor of the semiflexible polymer $S(k,N)$.
Performing the inverse Laplace transform by using Maple permits to get plots
of the structure factor.

In conclusion, we have studied the end-to-end distribution function of a
worm like chain, the problem which is equivalent to the Euclidean quantum
rigid rotator in a homogeneous external field, by using the propagator
method. We have established that the combinatorial problem of counting the
paths contributing to the end-to-end distribution function can be mapped to
the problem of random walks with constraints, which is closely related to
the Temperley-Lieb algebra. Using this mapping we derived an exact
expression for the Fourier-Laplace transform of the end-to-end polymer
distance as the matrix element of the inverse of an infinite rank matrix.
Starting with this result we have also derived a recursion relation
permitting to compute $G(k,p)$ directly. The mapping to the problem of
restricted random walks and the Temperley-Lieb algebra, which is not
restricted to the computation of $G(k,N)$, gives a new tool of treating the
semiflexible polymers. Eqs.(\ref{w6}-\ref{w7}) give also the exact solution
for the transition amplitude of a rigid quantum rotator in an external
homogeneous imaginary field in Euclidean\ sector.

\acknowledgments S.S. acknowledges a support from the Deutsche
Forschungsgemeinschaft (SFB 418 and Ste 981-1/1).

\end{document}